\documentclass[aps,prb,twocolumn]{revtex4}
\usepackage{bm}
\usepackage{graphicx}

\begin{document}

\title{Current fluctuations in composite conductors: Beyond the second cumulant}
\author{Artem V. Galaktionov$^{1}$ and Andrei D. Zaikin$^{2,1}$
}
\affiliation{$^1$I.E. Tamm Department of Theoretical Physics, P.N. Lebedev Physics Institute, 119991 Moscow, Russia}
\affiliation{$^2$Institute of Nanotechnology, Karlsruhe Institute of Technology (KIT), 76021 Karlsruhe, Germany}

\begin{abstract}
Employing the non-linear $\sigma$-model we analyze current fluctuations in coherent composite conductors which contain a diffusive element in-between two tunnel barriers. For such systems we explicitly evaluate the frequency-dependent third current cumulant which also determines the leading Coulomb interaction correction to
shot noise. Our predictions can be directly tested in future experiments.
\end{abstract}

\pacs{73.23.-b, 72.70.+m, 05.40.-a}

\maketitle

\section{Introduction}
Investigations of current fluctuations in mesoscopic conductors enable one to extract important information about their microscopic properties, such as, e.g.,  transmissions $T_k$ of their conducting channels. A great deal of information is
provided by the analysis of shot noise \cite{BB}. For instance, experimental confirmation of shot noise suppression in diffusive conductors (as compared to its Poisson value) by the Fano factor $\beta=1/3$ served as a strong argument supporting the validity of the universal transmission distribution $P(T_k)\propto 1/T_k\sqrt{1-T_k}$ in such conductors, see Refs. \onlinecite{BB,B} for more details on this subject.

It was also realized, that transmission distributions in composite conductors, such as, e.g., a tunnel junction attached to a diffusive wire can behave in a rather non-trivial way. For instance, it was observed experimentally that the low energy Andreev conductance of hybrid normal-superconducting structures may coincide with its conductance in the normal state. This zero-bias anomaly phenomenon \cite{LR} was explained \cite{BRM,N} by finding out that the transparency distribution for a composite system reduces to that of a diffusive conductor provided the resistance of the latter exceeds that of the tunnel barrier. This phenomenon can be interpreted as disorder-induced opening of tunneling channels.

Yet another non-trivial property of composite conductors involving many coherent scatterers is that shot noise in such systems turns out to be identical to that in a diffusive element even if none of these scatterers is diffusive. These property was initially established for chains of tunnel barriers \cite{JB} and later demonstrated also for other situations \cite{Sukh,GZ}. Most generally, one can consider a chain of $N$ different coherent conductors with arbitrary transmission distributions $T_k^{(n)}$ of their conducting channels and define their resistances $R_n$ and Fano factors $\beta_n$ as
\begin{equation}
\frac{1}{R_n}= \frac{2e^2}{h}\sum_k T_k^{(n)}, \quad \beta_n=\frac{\sum_k T_k^{(n)}(1-T_k^{(n)})}{\sum_k T_k^{(n)}}
\label{betan}
\end{equation}
with $e$ standing for the charge of the electron.
As usually, shot noise in such composite conductor with resistance $R=\sum_{n=1}^NR_n$ can be described by the correlation function
\begin{equation}
{\cal S}_2(\omega )=\int d\tau e^{i\omega\tau} \left\langle \delta I(t)\delta I(t-\tau ) \right\rangle ,
\label{shot}
\end{equation}
where $\delta I(t)=I(t)-\overline{I}$ accounts for current fluctuations around its average value $\overline{I}=V/R$ and $V$ is externally applied voltage bias. Here and below averaging $\langle ...\rangle$ should in generally be understood as the expectation value for the corresponding current operators. For the symmetric version of the correlator (\ref{shot}) in the zero frequency limit and at zero temperature one has ${\cal S}_2(0) =|e\overline{I}|\beta$, where the Fano factor $\beta$ of this general composite conductor reads \cite{GZ}
\begin{equation}
\beta
=\frac{1}{3}+\sum_{n=1}^N\frac{R_n^3}{R^3}\left(\beta_n-\frac{1}{3}\right).
\label{Farb}
\end{equation}
This formula clearly demonstrates that in the limit of sufficiently large $N$ one has $\beta \to 1/3$, i.e. a long chain of arbitrary -- not necessarily diffusive -- conductors universally behaves as a diffusive conductor.

Along with shot noise one can also investigate higher current cumulants. According to full counting statistics (FCS) theory \cite{LLS} these cumulants are in general expressed not only in terms of the Fano factor $\beta$ but also contain other combinations of channel transmissions. For instance, the third current cumulant
%\begin{multline}
\begin{eqnarray}{\cal S}_3(\omega_1,\omega_2)&=&\int d\tau_1 d\tau_2 e^{i\omega_1\tau_1+i\omega_2\tau_2}
\nonumber \\
&&\times \left\langle \delta I(t)\delta I(t-\tau_1) \delta I(t-\tau_2)\right\rangle
\label{third}
\end{eqnarray}
%\end{multline}
after its proper symmetrization \cite{GGZ1} can be cast to the form \cite{GGZ1,LevR}
\begin{equation}
{\cal S}_3(\omega_1,\omega_2)=\left(\beta-2\gamma F\right)e^2\overline{I},
\label{S3}
\end{equation}
where
\begin{equation}
\gamma=\frac{\sum_k T_k^2(1-T_k)}{\sum_k T_k}
\label{gamma}
\end{equation}
and $F$ is the function of bias voltage $V$, temperature $T$ and frequencies $\omega_{1,2}$ to be defined below. Thus, by studying the cumulant (\ref{third}) one
can recover not only the Fano factor $\beta$ but also the combination (\ref{gamma}). Experimental investigations of the third current cumulant have already started \cite{GabReul}, thereby making $\gamma$ (along with $\beta$) a directly measurable quantity. Furthermore, in the presence of electron-electron interactions the combination (\ref{gamma}) also affects the second current cumulant \cite{GGZ2}. Hence, $\gamma$ can in principle be extracted from
the noise measurements as well.

In the light of all these developments it would be useful to formulate a general approach enabling one to recover the parameter $\gamma$ (\ref{gamma}) for composite conductors. In the present paper we will address a specific situation that is frequently encountered in modern experiments embracing a variety of meso- and nanostructures: An arbitrary coherent diffusive conductor is connected to an external measuring scheme (leads) via tunneling interfaces. In other words, we
will consider a composite conductor  consisting of two tunnel barriers connected by a diffusive element. Employing the non-linear $\sigma$-model technique we will develop a method enabling one to evaluate the FCS generating function for this system. We will then derive the parameter $\gamma$ in terms of both tunneling and diffusive resistances involved in our problem. The corresponding expression can be directly used in future experiments investigating current fluctuations in such composite conductors. Conversely, information about three first current cumulants allows to fully determine the resistances of the three elements of the system under consideration.

Our paper is organized as follows. In Sec. II we work out a non-linear $\sigma$-model approach and establish general expressions for the parameters $\beta$ and $\gamma$ of our composite conductor. Our results and their possible extensions are then discussed in Sec. III. Some technical details of our calculation are relegated to Appendix.

 \section{Effective action and FCS}
We will consider a system which consists of a normal diffusive conductor with resistance $R_D$ and length $L$ attached to two big normal reservoirs via tunnel barriers (located at $x=0$ and $x=L$) with resistances $R_1$ and $R_2$ and cross-sections $\Gamma_1$ and $\Gamma_2$. This system is schematically depicted in Fig. 1.
We will assume that the whole system remains coherent, i.e. $L$ is shorter than both phase and energy relaxation lengths. We will also assume that the Thouless energy (inverse dwell time) of a composite conductor in-between two reservoirs is larger than any other energy scale in our problem.

\begin{figure}
\includegraphics[width=8cm]{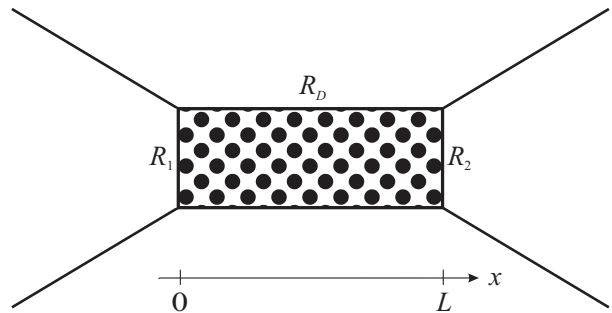}
\caption{Diffusive coherent conductor with resistance $R_D$ between two normal reservoirs attached via tunnel barriers with resistances $R_1$ and $R_2$. The barriers are located at $x=0$ and $x=L$. }
\end{figure}

In order to proceed below we will employ the Keldysh non-linear $\sigma$-model approach \cite{FN}.
The kernel of the evolution operator defined on the Keldysh contour can be expressed in terms of the path integral
\begin{equation}
\int {\cal D}\hat g {\cal D}\bm{A}{\cal D}V\exp (iS_{\rm fields}+iS),
\label{pathint}
\end{equation}
where $\bm{A}$ and $V$ are respectively the vector and scalar potential fields and
$\hat g (\bm{R}, t_1,t_2)$ are $2\times 2$ matrices which
depend on one coordinate and two time variables and obey
the normalization condition
\begin{equation}
\int dt'\hat g(\bm{R}, t_1,t') \hat g(\bm{R}, t',t_2)=\delta(t_1-t_2).
\label{norm}
\end{equation}
Here $S_{\rm fields}[\bm{A},V]$ is the action for the electromagnetic fields and
\begin{equation}
S=S_D+S_1+S_2
\label{saq1}
\end{equation}
defines the action for our composite conductor. Specifically,
\begin{equation}
S_D=\frac{i\pi N_0}{2}{\rm Tr}\,\left[ D\left( \bm
\partial \hat g \right)^2 + 4i\left(i
\partial_t -e\hat V\right)\hat
g\right], \label{saq2}
\end{equation}
is the non-linear $\sigma$-model effective action which accounts for a diffusive conductor and the terms
\begin{equation}
S_{1}=\frac{\pi \sigma_1}{2ie^2}{\rm Tr}_{\Gamma_{1}}\,\hat
g_1\hat g (0), \quad S_{2}=\frac{\pi \sigma_2}{2ie^2}{\rm Tr}_{\Gamma_{2}}\,\hat
g (L)\hat g_2
\label{saq3}
\end{equation}
describe tunnel barriers. Here and below  $\sigma_{1,2}=1/(R_{1,2} \Gamma_{1,2})$ and $\hat g_{1,2}$ represent the quasiclassical energy-integrated Green-Keldysh matrices of the reservoirs to be specified below. The expressions (\ref{saq3}) follow directly from Kupriyanov-Lukichev boundary conditions \cite{KupLuk}. These boundary conditions are sufficient to employ here since our further analysis will be restricted to the lowest order in barrier transmissions.

Here and below the product of matrices $\hat g$  should be
understood as a convolution, cf. Eq. (\ref{norm}). The trace is
taken over both space and time variables and it
is accompanied by a summation over matrix indices. The diffusion coefficient and the density of states per spin at the Fermi surface of a diffusive conductor are denoted by $D$ and $N_0$ respectively. Spatial integrations in the terms $S_{1,2}$ (\ref{saq3}) are restricted to the respective barrier interfaces.

The $2\times2$ matrix $\hat V$ term depends on the Hubbard-Stratonovich
fields $V^F$ and $V^B$ defined respectively on the forward and backward branches of the Keldysh contour. As usually, these fluctuating fields emerge after the standard
decoupling procedure in the Coulomb term in the Hamiltonian. Here we define
\begin{equation}
\hat V(\bm{R},t)=\left(\begin{array}{cc}V^+& \frac{1}{2}
V^-
\\ \frac{1}{2} V^- & V^+\end{array} \right),\label{vtot}
\end{equation}
where $V^+=(V^F+V^B)/2,\,
V^-=V^F-V^B$ are respectively the "classical" and "quantum"
fluctuating fields. Finally, we specify the operator
\begin{eqnarray}
&& \bm\partial \hat g= \nabla\hat g-i\frac{e}{c}
\left[\hat{\bm{A}},\hat  g\right]\equiv \label{partial}\\ &&
\nabla\hat  g(t,t') -i\frac{e}{c} \hat{\bm{A}}(t)\hat
g(t,t')+\hat g(t,t') i\frac{e}{c} \hat{\bm{A}}(t').
\nonumber
\end{eqnarray}
In Eq. (\ref{partial}) we suppressed spatial arguments for the sake of brevity, introduced the speed of light $c$ and defined the matrix
\begin{equation}
\hat{\bm{A}}=\left(\begin{array}{cc}\bm{A}^{+}& \frac{1}{2}
\bm{A}^{-}
\\ \frac{1}{2}
\bm{A}^- & \bm{A}^+.
\end{array} \right),
\end{equation}
which -- analogously to Eq. (\ref{vtot}) -- is composed of the classical $\bm{A}^+=(\bm{A}^F+\bm{A}^B)/2$ and quantum $\bm{A}^-=\bm{A}^F-\bm{A}^B$ components of the vector potential.

Now let us employ the above non-linear $\sigma$-model formalism in order to construct the FCS generating function. In what follows we will assume that an external dc voltage $V$ is applied to the system. This voltage is accounted for by a simple shift of the classical component $V^+ \to V+V^+$. At this stage we also neglect the effect of electron-electron interactions by setting the quantum component of the scalar potential equal to zero $V^-\to 0$. At the same time we will still keep the quantum component of the vector potential $\bm{A}^-$ which is needed for further evaluation of the current cumulants. Note that the frequency dependence of the third current cumulant in Eq. (\ref{S3}) is fully accounted for by the function $F$ evaluated in \cite{GGZ1}, while the parameters $\beta$ and $\gamma$ are frequency-independent. Hence, in order to determine these parameters it suffices to perform our calculation in the zero frequency limit. In this case, similarly to Ref. \onlinecite{KL} it is convenient to employ the pure gauge
\begin{equation}
\bm{A}^-=\nabla \chi .
\label{pg}
\end{equation}
We will see that $\chi$ is going to play the role of the counting field in the FCS generating function for our problem.

Now let us specify the Green-Keldysh matrices of the reservoirs. Setting the scalar potential of the left reservoir equal to $V$ we have in the frequency representation
\begin{equation}
\hat g_{1\epsilon}=\left( \begin{array}{cc} 1 & 2N(\epsilon -eV)\\ 0 &-1 \end{array} \right),\quad N(\epsilon)=\tanh
(\epsilon/2T). \label{gle}
\end{equation}
Accordingly, the scalar potential of the right reservoir equals to zero.
The counting field $\chi$ can be gauged away from the action (\ref{saq2}) at the expense
of the following transformation of the Green-Keldysh matrix of the right reservoir:
\begin{equation}
\hat g_{2\;\rm{new}}=\exp\left(-\frac{i\varphi}{2}\hat \sigma_x\right)\hat g_{2\;\rm{old}} \exp\left(\frac{i\varphi}{2}\hat \sigma_x\right).
\label{transf}
\end{equation}
where $\varphi=e\chi/c$ and
\begin{equation}
\hat\sigma_x=\left( \begin{array}{cc}0 &1\\  1& 0
\end{array}\right)
\end{equation}
is the Pauli matrix. The transformation (\ref{transf}) yields $\hat g_2$ in the form
\begin{eqnarray}
&& \hat g_{2\epsilon}=\label{gre}\\ &&  \left( \begin{array}{cc} \cos \varphi+i\sin \varphi N(\epsilon)&  i\sin \varphi+2\cos^2\frac{\varphi}{2}N(\epsilon)\\ - i\sin \varphi+2\sin^2\frac{\varphi}{2}N(\epsilon)& -\cos \varphi-i\sin \varphi N(\epsilon)
\end{array}\right).
\nonumber
\end{eqnarray}

What remains is to evaluate the path integral (\ref{pathint}) within the saddle point approximation. The saddle point paths obey the well known Usadel equation \cite{Usadel,BWBSZ}. In the stationary limit to be considered below this equation reads
\begin{equation}
D[\bm\partial ,\hat g_{\epsilon}(x)[\bm\partial ,\hat g_{\epsilon}(x)]]=[-i\epsilon,\hat g_{\epsilon}(x)].
\label{Usadel}
\end{equation}
Eq. (\ref{Usadel}) should be supplemented by Kupriyanov-Lukichev boundary conditions \cite{KupLuk} involving the Green-Keldysh matrices of the reservoirs (\ref{gle}) and (\ref{gre}). Once the solution of Eq. (\ref{Usadel}) is established, the generating function for current cumulants $iS(\chi)$ is immediately obtained by substituting this solution back into the action (\ref{saq1}).

Let us define the matrix
\begin{equation}
\hat J(\epsilon )=\hat g_{\epsilon} \partial_x \hat g_{\epsilon},
\end{equation}
which anticommutes with $\hat g_{\epsilon}$ and does not depend on the coordinate $x$. The latter property just reflects the spectral current conservation.  With the aid of this matrix the solution of Eq. (\ref{Usadel}) inside the diffusive conductor can be expressed in the form
\begin{equation}
\hat g_{\epsilon} (x)=\hat g_{\epsilon} (0)\exp\left( \hat J(\epsilon ) x\right).
\end{equation}
In addition, Kupriyanov-Lukichev boundary conditions yield
\begin{eqnarray}
&& \hat J(\epsilon )=\frac{\sigma_1}{2\sigma_D}\left[\hat g_{1\epsilon},\hat g_{\epsilon} (0)\right],\label{fe}\\
&& \hat J (\epsilon )=\frac{\sigma_2}{2\sigma_D}[\hat g_{\epsilon} (0) \exp(\hat J(\epsilon ) L), \hat g_{2\epsilon}],\label{se}
\end{eqnarray}
where $\sigma_D=2e^2 DN_0$  is the Drude conductivity.

Let us fix some parametrization of $\hat g_{\epsilon}(0)$ satisfying the normalization condition $\hat g_{\epsilon}^2(0)=1$. Then with the aid of Eq. (\ref{fe}) we evaluate the matrix $\hat J(\epsilon )$. Substituting the resulting expression into Eq. (\ref{se}) we determine the parameters employed in our parametrization for $\hat g_{\epsilon}(0)$. Further details of this calculation are presented in Appendix A. Here we evaluate the effective action up to the third order in $\chi$. Making use of the identity ${\rm Tr}\left( \partial_x\hat g\right)^2 =-{\rm Tr}{\hat J}^2$ and keeping the contribution from the interface terms we eventually arrive at the generating function
\begin{eqnarray}
&& iS(\chi )=\frac{i\overline{I}\chi}{c}-\frac{e\overline{I}\chi^2}{2c^2}\left( \beta\coth\frac{v}{2}+\frac{2(1-\beta)}{v} \right)\nonumber\\&& -\frac{ie^2\overline{I}\chi^3}{6c^3}\left(\beta-2\gamma F(v)\right)+{\cal O}(\chi^4), \label{fr}
\end{eqnarray}
where
\begin{eqnarray}
&& \beta=\frac{1}{3}+\frac{2}{3}\frac{(R_1^3+R_2^3)}{R^3},\label{bg}\\
&& \gamma=\frac{2}{15}+\frac{6}{5}\frac{(R_1^5+R_2^5)}{R^5}-\frac{4}{3
} \frac{(R_1^3+R_2^3)^2}{R^6}.\label{ga}
\end{eqnarray}
Here we defined $v=eV/T$, the total resistance $R=R_1+R_D+R_2$ and \cite{LevR}
\begin{equation}
F(v)=1+{3}\frac{1-\left(\sinh v/v \right)}{\cosh v-1}.\label{funcf}
\end{equation}
The generating function (\ref{fr})-(\ref{ga}) represents the central result of this paper. Making use of the correspondence $I \rightarrow -ic\delta/\delta\chi$
this result allows to recover the first three current cumulants for our composite conductor. It is important to emphasize that the expressions for these cumulants
-- including the third one \cite{GGZ1} -- can be fully described
not only in the zero frequency limit, but also at finite frequencies. This and some other applications of the above results will be briefly discussed in the next section.

\section{Discussion and outlook}
It is satisfactory to observe that our result for $\beta$ (\ref{bg}) is fully consistent with the general formula (\ref{Farb}) derived within a totally different framework. We also observe that, as compared to $\beta$, the expression for $\gamma$ (\ref{ga}) is determined by a substantially more complicated combination of resistances $R_1$, $R_2$ and $R_D$. Even more complicated expression for $\gamma$ can be expected in a general case of $N$ arbitrary conductors.

Let us specify our results for $\beta$ and $\gamma$ in
several important limits. Provided the resistance of one tunnel barrier dominates $R_1 \gg R_D,R_2$, Eqs. (\ref{bg}) and (\ref{ga}) yield $\beta \approx 1$ and $\gamma \approx 0$, i.e. in this limit current fluctuations are identical to those in a tunnel barrier. If, however, the diffusive resistance prevails $R_D\gg R_1,R_2$, we have $\beta \approx 1/3$ and $\gamma \approx 2/15$. These values are characteristic for any diffusive system. In the case
$R_1\approx R_2\gg R_D$ from Eqs. (\ref{bg}) and (\ref{ga}) we obtain
$\beta \approx 1/2$ and $\gamma \approx 1/8$. The latter value of $\gamma$ turns out to already be rather close to the diffusive one $\gamma =2/15$ rapidly approaching it with increasing $R_D$. E.g., for $R_1 \approx R_2 \approx R_D$ the Fano factor $\beta$ still remains bigger than $1/3$ by $\approx 15${\%} while $\gamma$ differs from 2/15 by less than 2{\%}.

Turning to the frequency dependence of the third current cumulant and making use
of the results \cite{GGZ1} we establish the function $F$ in Eq. (\ref{S3}) in the following general form
\begin{eqnarray}
&& F(v,w_1,w_2)=Z(v,-w_1,w_1+w_2)\label{fz}\\
&& +Z(v,-w_2,w_1+w_2) +Z(v,w_1,w_2), \nonumber
 \end{eqnarray}
where $w_{1,2}=\omega_{1,2}/2T$ and the function $Z$ reads
\begin{eqnarray}
&& Z(v,x,y)=\frac{\sinh(v/2)}{2v\sinh x\sinh y}\label{ZZZ}\\
&& \times\left(\frac{x+y-(v/2)}{\sinh[x+y-(v/2)]}  + \frac{x+y+(v/2)}{\sinh[x+y+(v/2)]} \right).
\nonumber
\end{eqnarray}
In the zero frequency limit $w_{1,2} \to 0$ the function $F$ (\ref{fz}) reduces to
that defined in Eq. (\ref{funcf}). The above equations together with Eqs. (\ref{S3}), (\ref{bg}) and (\ref{ga}) fully
describe the third current cumulant in our composite conductor provided its size is
restricted in such a way that its Thouless energy (inverse dwell time) remains the highest energy parameter in the problem. Note that the opposite physical limit of very long dwell times was studied for a chaotic cavity in Ref. \onlinecite{Nag}. The frequency dependence of ${\cal S}_3$ established in that limit \cite{Nag} differs from ours, however in the  limit of zero frequencies the results \cite{Nag,Nag1} are consistent with ours (taken at $T \to 0$) provided we set $R_D=0$ in Eqs. (\ref{bg}) and (\ref{ga}). We
also note that non-symmetric third cumulants of the current can also be analyzed \cite{Salo}. The corresponding results for ${\cal S}_3$ being in general different from Eq. (\ref{S3}) also contain the parameter $\gamma$ (\ref{gamma}) evaluated here.

Let us now take into account electron-electron interactions. It was argued \cite{GGZ2} that such interactions may affect shot noise. In particular, for $R \ll R_q$ and $T,|\omega|\ll |eV|\ll 1/RC$ we obtain \cite{GGZ2}
\begin{equation}
{\cal S}_2(\omega)=|e\overline{I}|\beta-\frac{2(\beta-2\gamma)|eV|}{R_q}
\ln\frac{1}{|eV|RC},\label{intc}
\end{equation}
where $R_q=h/e^2$ and $C$ are respectively the quantum resistance unit and the effective sample capacitance. This result also follows directly from the renormalization group
analysis \cite{KBN,GGZ3} stating that electron-electron interactions effectively yield energy dependent renormalization of channel transmissions.
Eq. (\ref{intc}) demonstrates that the magnitude of the leading interaction correction to the second current cumulant ${\cal S}_2$ is governed by
the combination $\beta -2\gamma$. This combination -- depending on the system -- can take either positive or negative values. Accordingly, electron-electron interactions can either suppress or enhance shot noise. In the situation considered here the parameter $\beta -2\gamma$ is always positive ranging from 1/15 to 1 depending on the resistance values. Hence, in our case Coulomb interaction always tends to suppress shot noise.

It is also worth pointing out that our results (\ref{bg}) and (\ref{ga}) can be rewritten in terms of the following equations
\begin{eqnarray}
&&\left( \frac{R_1}{R}\right)^3+\left( \frac{R_2}{R}\right)^3=\frac{3\beta-1}{2}, \\ && \left( \frac{R_1}{R}\right)^5+\left( \frac{R_2}{R}\right)^5=\frac{15\gamma+5(3\beta-1)^2-2}{18}.\nonumber
\end{eqnarray}
Two real solutions of these equations (corresponding to $R_1\leftrightarrow R_2$) can be easily found numerically. Thus, information about the first three current cumulants is sufficient to determine all three resistances $R_1$, $R_D$ and
$R_2$ in the system under consideration.

Finally, we would like to note that our analysis also allows to derive the fourth and even higher current cumulants for our composite conductor. For this purpose it is
necessary to establish higher order in $\chi$ terms in the expression for the generating function $iS(\chi )$. This calculation, although quite tedious, can be performed in a straightforward manner along the same lines as it was demonstrated here. Yet another promising extension of our formalism could be to apply it to hybrid normal-superconducting structures. In this case modifications simply amount to including superconductivity into the Usadel equations in the standard manner. The corresponding analysis, however, is beyond the frames of the present paper.

\vspace{0.5cm}

\centerline{\bf Acknowledgements}

\vspace{0.5cm}

This work was supported in part by RFBR under grant 09-02-00886.

\appendix
\section{}
Let us present some details of our calculation. The matrix $\hat g_{\epsilon}(0)$ satisfying the normalization condition $\hat g^2_{\epsilon}(0)=1$ can be written up to the third order in $\chi$ as
\begin{eqnarray}
&& \hat g_{\epsilon}(0)=\left( \begin{array}{cc} 1& 2f\\ 0& -1\end{array}\right)+ \left( \begin{array}{cc} -fp_3& p_2\\p_3 & f p_3\end{array}\right) \nonumber +
\\ && \left( \begin{array}{cc} -fq_3-\frac{1}{2}\left(f^2p_3^2+p_2p_3\right)& q_2\\q_3 & f q_3+\frac{1}{2}\left(f^2p_3^2+p_2p_3\right)\end{array}\right)\nonumber\\
&& +\left( \begin{array}{cc} -fr_3-B& r_2\\r_3 & f r_3+B\end{array}\right),
\end{eqnarray}
where
\begin{equation}
B=f^2 p_3q_3+\frac{1}{2}\left(fp_2p_3^2+f^3p_3^3+p_3q_2+p_2q_3 \right).
\end{equation}
Resolving the self-consistency equations (\ref{fe}) and (\ref{se}), for $f$ and $p_3$ we obtain
\begin{eqnarray}
f=\frac{R_1}{R}N(\epsilon)+\frac{R_2+R_D}{R}N(\epsilon -eV),\;
 p_3=-\frac{i\varphi R_1}{R}.\label{fochi}
\end{eqnarray}
The expressions (\ref{fochi}) account for the first order in $\chi$ term in the generating function $iS(\chi)$. The contributions $i\overline{I}\chi R_{1,2}/c R  $ come from the interface terms (\ref{saq3}), while the term $i\overline{I}\chi R_D/c R  $ emerges from Eq. (\ref{saq2}) describing a diffusive element conductor. Their sum yields the first term in Eq. (\ref{fr}), i.e. just the Ohm's law for our composite conductor.

Next let us reconstruct the second order in $\chi$ contribution to the generating function $iS(\chi )$. For that purpose we need to determine the parameters $p_2$ and $q_3$. They are
\begin{widetext}
\begin{eqnarray}
&& p_2(\epsilon)=\frac{2i\varphi R_1}{3R}\left[ N^2(\epsilon -eV)\left(2-\frac{3 R_1}{R}+\frac{R_1^3+R_2^3}{R^3} \right)-N^2(\epsilon)\left(1-\frac{R_1^3+R_2^3}{R^3} \right) \right. \nonumber\\ &&
\left. -N(\epsilon -eV)N(\epsilon)\left(1-\frac{3 R_1}{R}+\frac{2(R_1^3+R_2^3)}{R^3} \right) \right]
+\frac{i\varphi R_1}{R},\\
&& q_3(\epsilon)=\frac{R_1\varphi^2}{2R}\left[ N(\epsilon-eV)+\frac{1}{3}\left( N(\epsilon)-N(\epsilon-eV) \right)\left(1+\frac{2\left(R_1^3+R_2^3\right)}{R^3}  \right)\right].\nonumber
\end{eqnarray}
\end{widetext}
The zero-frequency current noise follows from the second term in Eq. (\ref{fr}).

Finally, in order to find the third current cumulant it is necessary to find the parameters $q_2$ and $r_3$. The corresponding calculation is straightforward but yields rather lengthy expressions which we do not present here.  One encounters the following integrals
\begin{eqnarray}
&& \int\limits_{-\infty}^\infty dx\left( \tanh x-\tanh(x-a)\right) \tanh x\tanh(x-a) \nonumber\\
&&=\frac{2 a}{3}-\frac{8a}{3}F(2a),\\
&&  \int\limits_{-\infty}^\infty dx\left( \tanh x-\tanh(x-a)\right) \tanh^2 x
\nonumber\\ &&=\frac{2 a}{3}+\frac{4a}{3}F(2a),\nonumber
\end{eqnarray}
which contain the function $F(v)$ (\ref{funcf}). Collecting all terms one eventually arrives at the third order in $\chi$ contribution to the generating function (\ref{fr})-(\ref{ga}).


\begin{thebibliography}{}
\bibitem{BB} Ya.M. Blanter and M. B\"uttiker, Phys. Rep. {\bf 336}, 1 (2000).
\bibitem{B} C.W.J. Beenakker, Rev. Mod. Phys. {\bf 69}, 731 (1997).
\bibitem{LR} C.J. Lambert and R. Raimondi, J. Phys.: Condens. Matter {\bf 10}, 901 (1998).
\bibitem{BRM} C.W.J. Beenakker, B. Rejaei, and J.A. Melsen, Phys. Rev. Lett. {\bf 72}, 2470 (1994).
 \bibitem{N} Yu.V. Nazarov, Phys. Rev. Lett. {\bf 73}, 134 (1994).
\bibitem{JB} J.M. de Jong and C.W.J. Beenakker, Phys. Rev. B {\bf 51},
16867 (1995).
\bibitem{Sukh} S. Oberholzer, E.V. Sukhorukov, C. Strunk, and C.
Sch\"onenberger, Phys. Rev. B {\bf 66}, 233304 (2002).
\bibitem{GZ} D.S. Golubev and A.D. Zaikin Phys. Rev. B {\bf 70}, 165423 (2004).
\bibitem{LLS} L.S. Levitov, H.W. Lee, and G.B. Lesovik, J. Math. Phys. {\bf 37}, 4845 (1996).
\bibitem{GGZ1} A.V. Galaktionov, D.S. Golubev, and A.D. Zaikin,  Phys. Rev. B {\bf 68}, 235333 (2003).
\bibitem{LevR} L.S. Levitov and M. Reznikov, Phys. Rev. B {\bf 70}, 115305 (2004).
\bibitem{GabReul} J. Gabelli and B. Reulet, J. Stat. Mech. P01049 (2009).
\bibitem{GGZ2} 	A.V. Galaktionov, D.S. Golubev, and A.D. Zaikin, Phys. Rev. B {\bf 68}, 085317 (2003).
\bibitem{FN} An alternative (and equivalent) way would be to make use of DMPK equations \cite{B}.
\bibitem{KupLuk} M.Yu. Kupriyanov and V.F. Lukichev, Zh. Eksp. Teor. Fiz. {\bf 94}, 139 (1988) [Sov. Phys. JETP {\bf 67}, 1163 (1988)].
\bibitem{KL} A. Kamenev and A. Levchenko, Adv. Phys. {\bf 58}, 197 (2009).
\bibitem{Usadel} K.D. Usadel,  Phys. Rev. Lett. {\bf 25}, 507 (1970).
\bibitem{BWBSZ} W. Belzig, F.K. Wilhelm, C. Bruder, G. Sch\"on, and A.D. Zaikin, Superlatt. Microstruct. {\bf 25}, 1251 (1999).
\bibitem{Nag} K.E. Nagaev, S. Pilgram, and M. B\"uttiker,
Phys. Rev. Lett. {\bf 92}, 176804 (2004).
\bibitem{Nag1} K.E. Nagaev, P. Samuelsson, and S. Pilgram, Phys. Rev. B {\bf 66}, 195318 (2002).
\bibitem{Salo} J. Salo, F.W.J. Hekking, and J.P. Pekola, Phys. Rev. B {\bf 74}, 125427 (2006).
\bibitem{KBN} M. Kindermann and Yu.V. Nazarov, Phys. Rev. Lett. {\bf 91}, 136802 (2003); D.A. Bagrets and Yu.V. Nazarov, Phys. Rev. Lett. {\bf 94}, 056801 (2005).
\bibitem{GGZ3} D.S. Golubev and A.D. Zaikin, Phys. Rev. B {\bf 69}, 075318 (2004); D.S. Golubev, A.V. Galaktionov, and A.D. Zaikin, Phys. Rev. B {\bf 72}, 205417 (2005).

\end{thebibliography}
\end{document}